\documentclass[10pt]{article}
\usepackage{tikz}
\usepackage{xcolor}
\usepackage{subfig}
\usepackage{listings} 
\usepackage{booktabs}
\usepackage[utf8]{inputenc}
\usepackage{braket}
\usepackage{algpseudocode} 
\usepackage{algorithm}
\usepackage{url}
\usepackage[bookmarks=true,breaklinks=true,letterpaper=true,colorlinks,citecolor=blue,linkcolor=blue,urlcolor=blue]{hyperref}

\usepackage{amssymb}
\usepackage{amsmath}

\newcommand{\toolname}{CircInspect}

\usepackage[backend=biber, url=false]{biblatex}
\title{\toolname: Integrating Visual Circuit Analysis, Abstraction, and Real-Time Development in Quantum Debugging}

\author{{Mushahid Khan, Prashant J. Nair, Olivia Di Matteo}\\
\textit{Electrical and Computer Engineering} \\
\textit{The University of British Columbia}\\
Vancouver, Canada \\
mkhan103@student.ubc.ca, \{prashantnair, olivia\}@ece.ubc.ca
}

\begin{document}

\maketitle

\begin{abstract}
    Software bugs typically result from errors in specifications or code translation. While classical software engineering has evolved with various tools and methodologies to tackle such bugs, the emergence of quantum computing presents unique challenges. 
    Quantum software development introduces complexities due to the probabilistic nature of quantum computing, distinct algorithmic primitives, and potential hardware noise.  In this paper, we introduce \toolname, an interactive tool tailored for debugging quantum programs in Python and PennyLane~\cite{pennylane}. By leveraging breakpoints and real-time software development features, \toolname~empowers users to analyze isolated quantum circuit components, monitor program output, visualize structural changes, and abstract information to enhance comprehension.
\end{abstract}


%

\section{Introduction}
%
%
%
%
 Software engineering involves designing, developing, testing, and maintaining software. It advances the development of software products through a structured software development cycle (SDC) that includes requirement analysis, system design, implementation, testing, debugging, deployment, and maintenance~\cite{bassil2012simulation}. However, existing classical software development models fall short for quantum software development~\cite{qcsdlc1} because of its distinct behavioral and computational patterns, limitations of early-stage quantum devices, and various implementation challenges.
Instead, proposals for an SDC tailored to a quantum setting have been put forward~\cite{weder2020quantum, qcsdlc1, qcsdlc2}.

Debugging is fundamental to both classical and quantum SDCs. A software bug arises from specification or code translation mistakes. Classical software debugging has evolved with diverse tools and methodologies ensuring system reliability~\cite{perscheid2017studying, hailpern2002software, zhou2004iwatcher}. In classical SDCs, common bugs include logical errors and workflow disruptions. Logic errors may not halt a program or generate error messages. However, they will cause programs to behave unexpectedly. Similarly, workflow issues arise from sequencing problems in the software's operations. Although such bugs can also appear in quantum programs, quantum computing has its own unique set of challenges. 

Quantum bugs arise due to the unique nature of quantum algorithms and thus require domain-specific knowledge to fix~\cite{di2024need}. Several studies have aimed to comprehensively grasp the nature of quantum bugs, including their characteristics, occurrence patterns, manifestations across diverse programming languages and frameworks, and strategies for prevention. However, creating effective quantum debugging tools requires interdisciplinary expertise in quantum physics, mathematics, computer science, and software engineering. This makes developing robust quantum software tools a demanding and ongoing endeavor.

Broadly, an effective quantum debugging tool must satisfy several crucial requirements:
\begin{enumerate}
\item[R1]It should enable visualization at various levels of abstraction within the quantum program. It enables users to delve into specific code segments without overwhelming the interface with unnecessary details~\cite{wen2023quantivine}.
  \item[R2] It should display the output of the quantum program at various execution stages, aiding users in swiftly pinpointing unexpected outcomes (quantum-specific bugs often appear as unexpected outputs~\cite{paltenghi2022bugs, luo2022comprehensive, camara2022fine, zhao2021bugs4q}).
  

  \item[R3]It should selectively display executed functions and their inputs and quantum operations. These provide a clear view of program execution and help users understand the program's behavior. 
  \item[R4]It must respect control flow, including conditionals, ensuring that debugging processes align with the logical flow of the program for accurate analysis and troubleshooting. This is key for mid-circuit measurements~\cite{midcircuitmeasurement} and post-selection in quantum algorithms.
\end{enumerate}

To that end, our paper introduces \toolname, an interactive tool for debugging quantum programs written in Python and PennyLane. \toolname~visualizes quantum programs as circuits at different levels of abstraction. It enables developers to visually isolate and examine quantum circuit components and monitor changes in program structure and output at breakpoints. It allows them to selectively observe inputs to subroutines and abstract away information for improved program comprehension. Additionally, \toolname~offers a real-time visualization mode that updates the quantum circuit dynamically as the user types their code.

This paper begins with a review of existing methodologies and tools for debugging in quantum computing, followed by an in-depth exploration of \toolname. We analyze \toolname's two modes of operation: the debugger mode's visual quantum circuit analysis and function abstraction, and the real-time development mode's monitoring of program changes and output. We then outline our evaluation plans and upcoming work.

\section{Background}
\subsection{Existing Work - Quantum Bugs}
Previous research has demonstrated the existence of diverse types of bugs that can have detrimental effects on quantum programs, emphasizing the need for specialized tools to address them. Analyzing and categorizing bugs aids in understanding their origins and identifying preventive measures. Bugs4Q \cite{zhao2021bugs4q} collected bugs within Qiskit \cite{aleksandrowicz2019qiskit} components (Terra, Aer, Ignis, and Aqua) and provided test cases for replicating erroneous behaviors. Bugs encountered include wrong output, and programs not implementing specific functions. For these, tools that provide features such as tracking the output throughout the quantum program and visualization to show which functions are implemented and which are not can help. \cite{zhao2021identifying} identified bug patterns in Qiskit and proposed remedies and preventative strategies. A tool to narrow down the subroutines in which these patterns occur can help users find and eventually deal with them. Prior work~\cite{zhao2023empirical} explored bugs in quantum machine learning frameworks, finding that 28\% of these bugs were quantum-specific, mainly involving qubit manipulation and function errors. They noted that issues like framework version mismatches and device compatibility were common, while algorithmic and logical flaws constituted a significant portion of the bugs. It is important to see how the algorithm output changes over time. This can help identify algorithmic and logical flaws. 

\subsection{Existing Work - Quantum Debugging Tools}
Building a quantum debugging tool is a non-trivial and complex task. It requires an in-depth knowledge of quantum mechanics and software engineering principles. Work has been done to develop both graphical and code-based tools. An effective tool should provide user-friendly information, offering features such as setting breakpoints, real-time visualization, tracking executed functions, and support for code import.

\subsubsection{Code-Based Quantum Debugging}

As a first line of defense, classical software engineering debugging techniques such as backtracking, cause elimination, and brute force debugging can be applied in the context of quantum debugging \cite{miranskyy2020your, miranskyy2021testing}. Following this, one can employ code-based quantum debugging tools, i.e., tools without a dedicated visual user interface. \cite{huang2019statistical} presents quantum program assertions based on statistical tests, while \cite{li2019proq} proposes projection-based runtime assertions. Cirquo \cite{metwalli2024testing} is a unit testing package developed using Qiskit and Python, featuring a slicer to segment a quantum circuit into subcircuits for categorization and testing. QChecker \cite{zhao2023qchecker} detects bugs in Qiskit code, informed by common bug patterns in quantum programs.

While these tools provide valuable insights into quantum programs, they are not standalone solutions. Users are required to integrate additional elements into their quantum programs, such as quantum operations or custom Python packages. Consequently, users must engage in additional quantum programming tasks for debugging, which could inadvertently introduce more bugs. Ideally, users should have access to a tool that can be invoked with minimal user overhead.

\subsubsection{Visual Quantum Debugging}
 Visualization can help us understand the structure and layout of quantum programs. In the absence of higher-level programming constructs, quantum circuits serve as a common visual aid for research, teaching, and algorithm development. While circuits sit at a very low level, they can be depicted with varying levels of abstraction. Many quantum programming frameworks offer tools for displaying circuits as either images or text. Some frameworks and tools extend this capability further by providing a more interactive interface. 

Quantivine \cite{wen2023quantivine} is an interactive tool that takes as input a Qiskit program and provides a series of visual representations to facilitate the comprehension and analysis of its circuit representation. 
This is done by processing the Qiskit code with circuit compilation, semantic analysis, and data alignment to extract meta-data and semantic information from the circuit.
These features allow users to explore and analyze quantum circuits interactively. Although Quantivine is not designed for debugging, the authors discussed a usage scenario where it can be used to observe gates in subroutines, aiding in bug identification. As such, it has limitations that don't allow for more in-depth debugging tasks. It lacks real-time visualization capabilities to depict quantum circuits that may change during execution; if a function includes conditionals, Quantivine does not visualize any gates within that function. Additionally, no error messages are provided for incorrect usage of Qiskit.

Drag-and-drop circuit simulators are also common.  Quirk \cite{quirk} is a drag-and-drop tool that operates in the browser. It is ideal for exploring the behavior of small quantum circuits and allows real-time diagnostics on a quantum circuit. However, Quirk does not support importing code. The IBM Quantum Composer \cite{composer} is a graphical quantum programming tool with drag-and-drop features and OpenQASM support, which can be used to program circuits and run them on real quantum hardware or simulators. Like Quirk, it is suitable for performing real-time diagnostics on small quantum circuits, but scalability is an issue with larger ones. In both tools, visualizations are limited to the granularity of individual gates.

Qiskit’s Trebugger \cite{Timeline} is a debugger for the transpiler. It inspects quantum circuits before and after transpilation and can identify specific phases in the transpilation process that may be responsible for producing unexpected output. Trebugger requires the qiskit\_trebugger package \cite{trebeggerpackage}.  It provides two views: cli, which is the default for terminal use, and jupyter, which is recommended for interactive and detailed transpilation viewing in a Jupyter notebook.

Q\# \cite{svore2018q} is an open-source programming language for quantum algorithm development. A Visual Studio extension for the Azure Quantum Development Kit incorporates a debugger designed specifically for Q\# programs. This debugger allows users to set breakpoints, navigate through their code step by step, delve into individual functions or operations, and monitor local variables, the quantum state of qubits, and circuit visualization. However, Q\#'s circuit visualization does not abstract different components of the quantum program; instead, it displays each executed quantum gate on a single circuit. Additionally, Q\# lacks a real-time development component.

Classiq \cite{classiq} is a platform that synthesizes a quantum circuit from a high-level functional model. It abstracts individual quantum operations into function blocks on the circuit, enabling users to expand these blocks and examine the underlying quantum operations. However, Classiq primarily serves as a visualization tool for the final quantum circuit, lacking features such as breakpoints to display the circuit's development over time or a real-time development component.

\section{\toolname}
Having explored the existing efforts in quantum debugger development and understanding the challenges they entail, we present \toolname. \toolname~is an interactive visual tool for debugging quantum circuits written in Python and PennyLane. \toolname~focuses on simulator-based debugging exclusively. It has two modes: debugger mode and real-time development mode. In debugger mode, users can debug quantum programs using a visual circuit and a tree structure of functions. On the other hand, real-time development mode allows users to observe their quantum program's circuit and output in real time. The methodologies we employ have the potential to be expanded to encompass other high-level quantum programming languages and frameworks. 

\subsection{Technology Stack and Development Approach}
\subsubsection{PennyLane}
Our research is centered around PennyLane because its programming paradigm aligns with our requirements for an effective quantum debugger. PennyLane facilitates programming quantum functions using subroutines and transforms, enabling function abstraction. Abstraction is critical for a debugger, allowing users to focus on specific code segments without being overwhelmed by irrelevant details (R1). Furthermore, PennyLane supports abstraction in circuit visualization, providing a clear and intuitive representation of program structure. It also enables the execution of quantum programs, including conditional quantum operations based on parameters and mid-circuit measurements (R2 and R4).

\begin{figure}[htbp]
\centerline{ \includegraphics[scale=0.5]{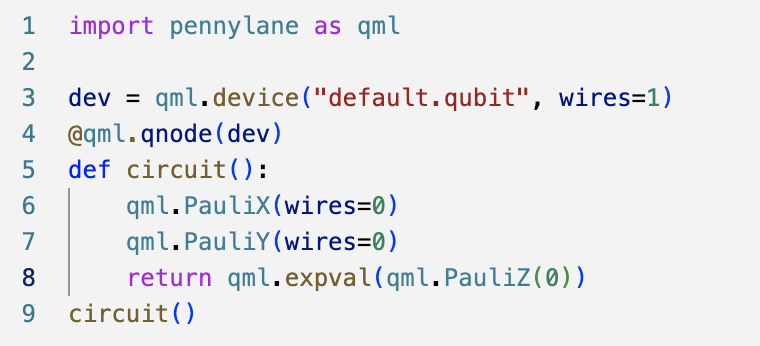}}
\caption{The QNode \texttt{circuit}, called in line 9, applies two operations and measures the expectation value of Pauli Z.}
\label{fig:pennylanecode}
\end{figure}

In PennyLane, quantum programs are expressed as \emph{quantum functions}. Quantum functions are Python functions that apply quantum operations and return one or more quantum measurements. To execute a quantum function, we require a device that can be either a simulator or actual hardware. A quantum function and device are bound together to create a quantum node (QNode), which can then be executed. \autoref{fig:pennylanecode} illustrates an example PennyLane program. In line 3, PennyLane's standard qubit-based device is created. For \toolname, QNodes must be created by decorating a quantum function with the \texttt{@qml.qnode} decorator. A decorator is a design pattern in Python that adds new functionality to an existing object without modifying its structure; in line 4, it promotes the quantum function, \texttt{circuit} to a QNode. The QNode must be invoked to execute this program, as shown in line 9.
\subsubsection{Front-end and Back-end Development Technologies}
\toolname~leverages React \cite{react} for the front-end, ensuring a user-friendly and interactive interface. For the backend, Python in conjunction with Flask \cite{flask} is used. Building \toolname~from scratch offers advantages over developing it as an extension in an integrated development environment (IDE). We, as developers, have complete control over the design, functionality, and customization of \toolname~to satisfy the requirements mentioned in the introduction. This approach allows for greater flexibility in choosing frameworks, ensuring compatibility with evolving quantum computing platforms and methodologies. Additionally, building independently facilitates updates, maintenance, and scalability without being limited by the constraints or dependencies of an IDE. When released, \toolname~will be hosted online so users can just go to a web page without installing specific tools.

\begin{figure*}[h!]
    \centerline
         {\includegraphics[width=\linewidth]{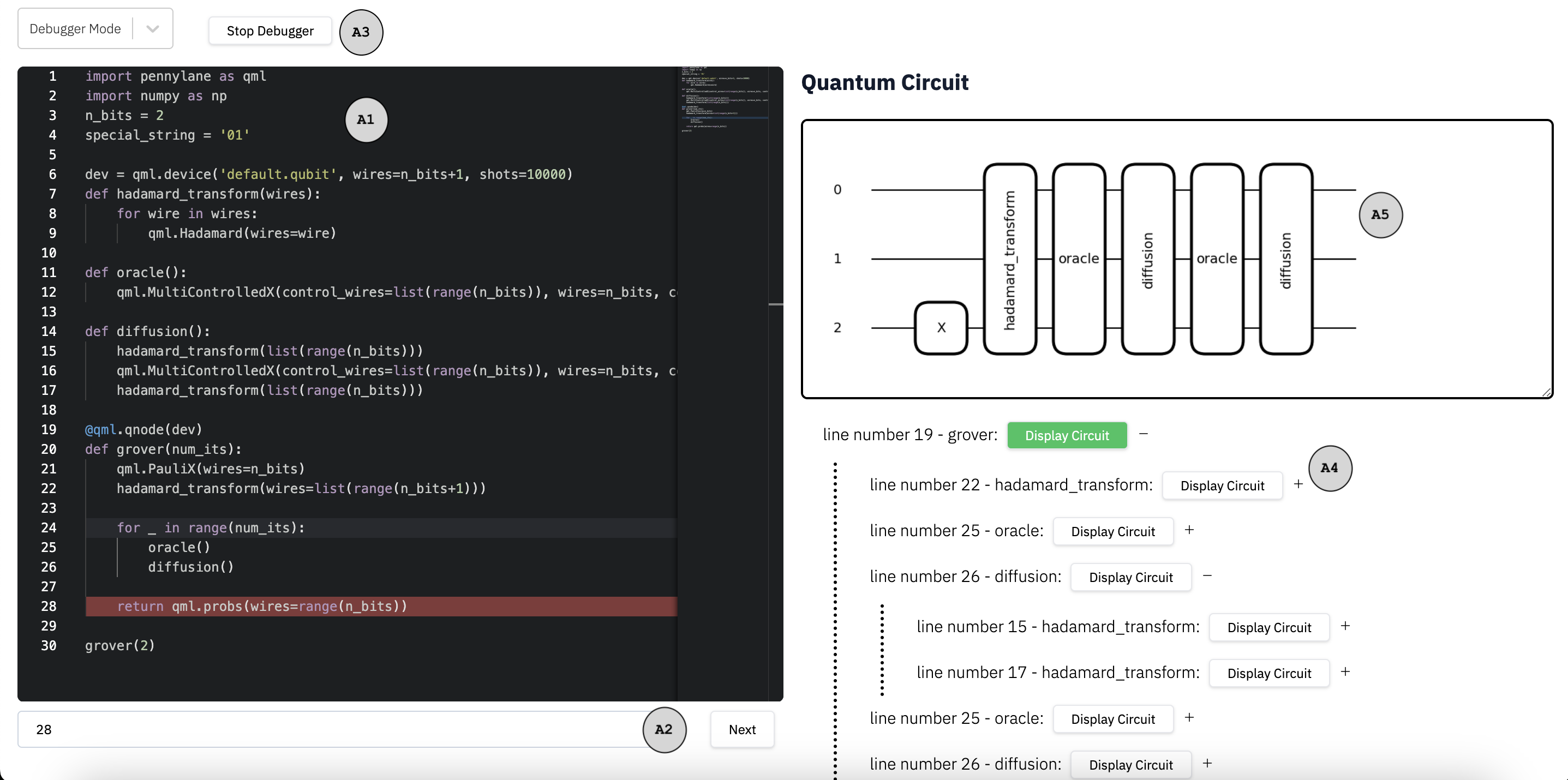}}
        \caption{Debugger mode being used for Grover's algorithm. The user inputs code in A1, sets breakpoints in A2, and hits the `Start Debugger' button A3. If there are syntax errors in A1, they are displayed in A4. Otherwise, A4 displays the structure of function calls and errors, and A5 visualizes the QNode's quantum circuit with function calls abstracted.}
        \label{fig:debugger_grover_main}
\end{figure*}


\subsection{Debugger mode}
\subsubsection{Using the Debugger mode} The debugger mode can be seen in \autoref{fig:debugger_grover_main}. A user uploads PennyLane code in Section A1, enters breakpoint line numbers in Section A2, and clicks the `Start Debugger' button (A3). Section A2 then updates to include a `Next' button. If there are syntax errors, they are displayed in A4. Otherwise, the user clicks `Next' and code execution halts at the next breakpoint. Each breakpoint triggers updates to the tree structure (A4) and circuit diagram (A5). 
\subsubsection{Benefits of the Debugger Mode}
Consider Grover's algorithm \cite{jozsa1999searching}, shown in Algorithm~\ref{alg:grover}.
\begin{algorithm}
            \textbf{Input:} \\
            \qquad$\bullet$  An oracle operator: $\ket{x}\ket{q}\rightarrow\ket{x}\ket{q\oplus f(x)}$\\
\qquad$\bullet$ $n$ + 1  qubits, $n$ is the length of each binary string\\
            \textbf{Output:}\\
\qquad$\bullet$ The unique bit string x satisfying $f(x) = 1$\\
	\caption{Grover's Algorithm} 
   \textbf{Procedure:}
	\begin{algorithmic}
            \State 1) Apply Pauli $X$ on the last qubit
            \State 2) Hadamard transform on all qubits
        \State 3) Repeat approximately $\sqrt(2^n)$ times:
				\State\hspace{0.2cm} 3.1) Apply oracle operator
    				\State\hspace{0.2cm} 3.2) Apply diffusion operator
	\State 4) Measure the first $n$ qubits and obtain the string $x$ that satisfies $f(x) = 1$ with high probability
	\end{algorithmic} 
 \label{alg:grover}
\end{algorithm}
When implementing Grover's algorithm, it is beneficial to encapsulate the Hadamard transform, oracle, and diffusion operator into separate functions. \autoref{fig:debugger_grover_main} showcases the debugger in use for Grover with a breakpoint set to line 28. After clicking `Next', execution halts, as highlighted in line 28.

\begin{figure}[ht]
    \centerline
         {\includegraphics[width=\linewidth]{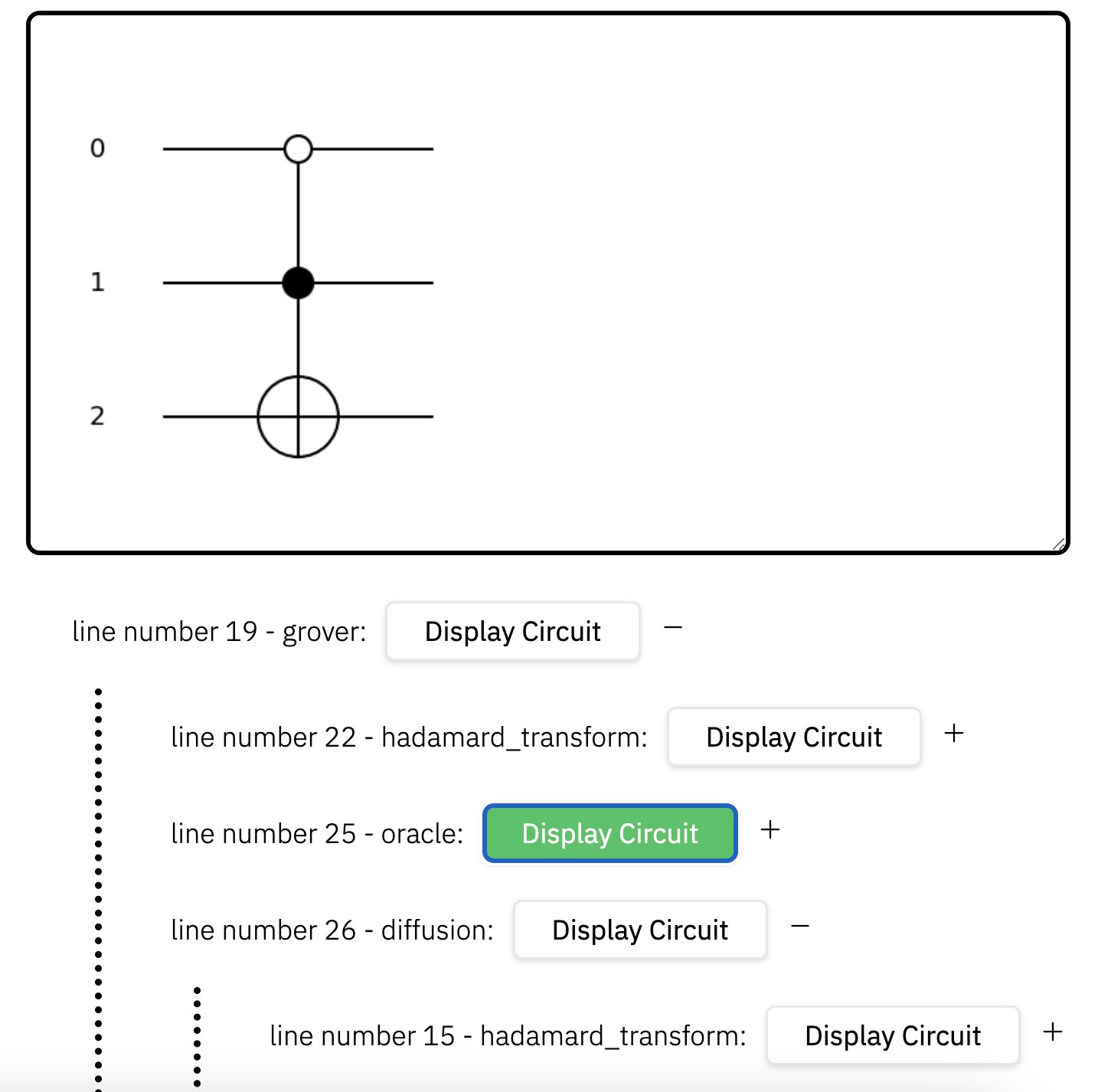}}
        \caption{The oracle function circuit is displayed.}
        \label{fig:sub_circuit}
\end{figure}

The visualization of the quantum circuit (A5) mirrors the subroutine structure in the quantum function \texttt{grover}. The circuit shows individual quantum operations within the main function but abstracts subroutines. While visualizing all gates in a single circuit may be manageable for short circuits, practical algorithms entail numerous nested subroutines, making it challenging to comprehend an algorithm's behavior through inspection at the level of individual gates. To avoid this, in \toolname, users can zoom into any subroutine by clicking  `Display Circuit'  next to the function name in A4. This action updates the circuit diagram to show that the subroutine's circuit with any nested function calls abstracted. An example of the Oracle function is shown in \ref{fig:sub_circuit}.

Abstraction facilitates the detection and elimination of erroneous quantum operations by providing a view of quantum operations within functions and their execution in the code. This capability assists in pinpointing redundant code segments. Put simply, users can examine the abstracted quantum circuit to compare operations within subroutines against those in the code, distinguishing between executed and unused lines. Consequently, users can efficiently remove unnecessary code lines.
The abstraction also comes in handy when a subroutine's structure can change due to input parameters. If a subroutine has a different structure each time it is called, it is harder to tell which subroutine call a gate belongs to if abstraction is not present. This can happen if there are conditionals present in a subroutine.

\begin{figure}[h!]
    \centerline
         {\includegraphics[width=\linewidth]{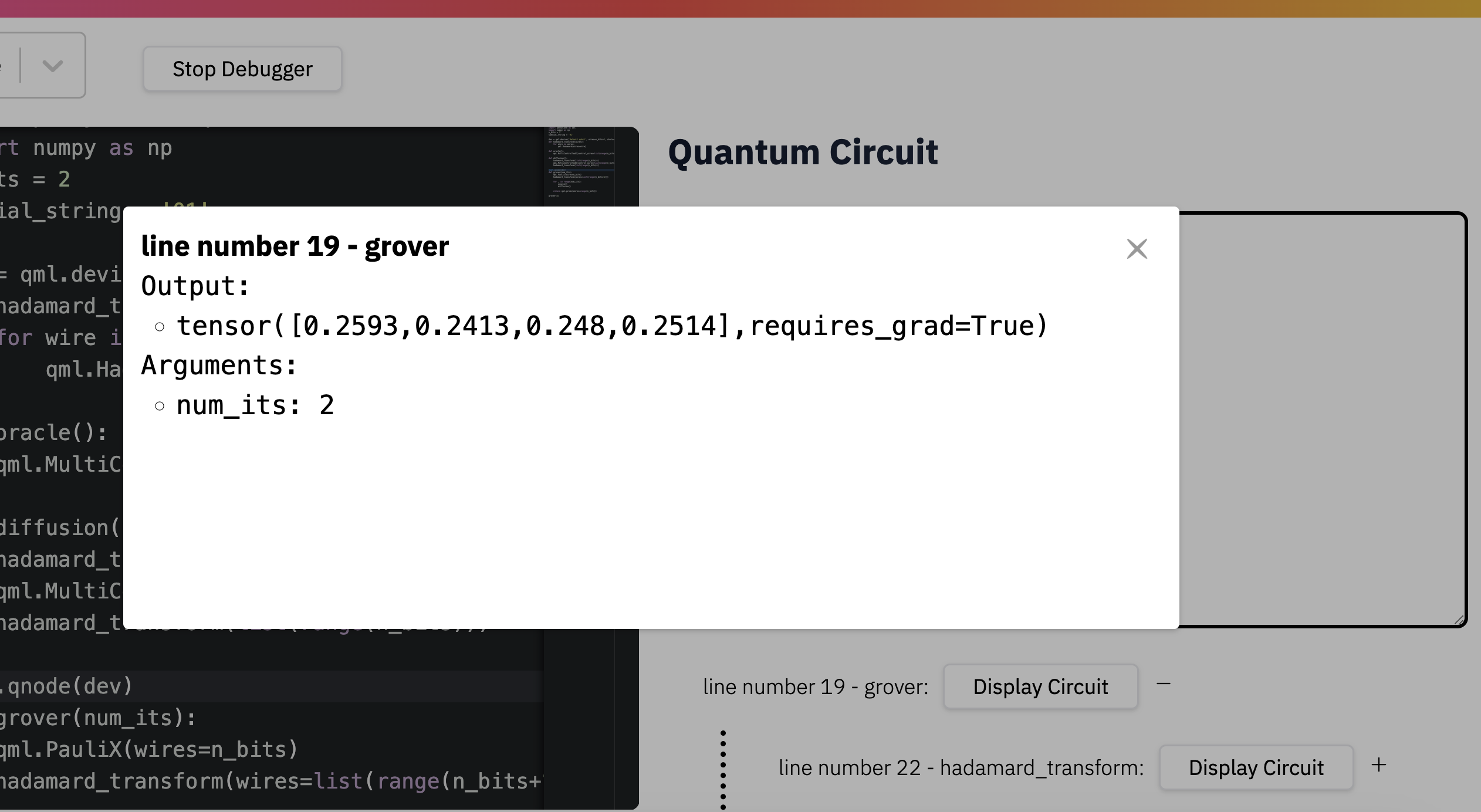}}
        \caption{The main function's output and the arguments for any function at a specific breakpoint.}
        \label{fig:fcn_output}
\end{figure}

The tree structure (A4) facilitates the swift identification of invoked functions. Users can also determine the main function's output and the arguments for any function at a specific breakpoint by clicking the function name, as seen in \autoref{fig:fcn_output}. This is done on an as-needed basis to keep the visualization clean. It lets users pinpoint unexpected output and detect incorrect arguments passed to functions.

\begin{figure*}[h!]
    \centerline
         {\includegraphics[width=\linewidth]{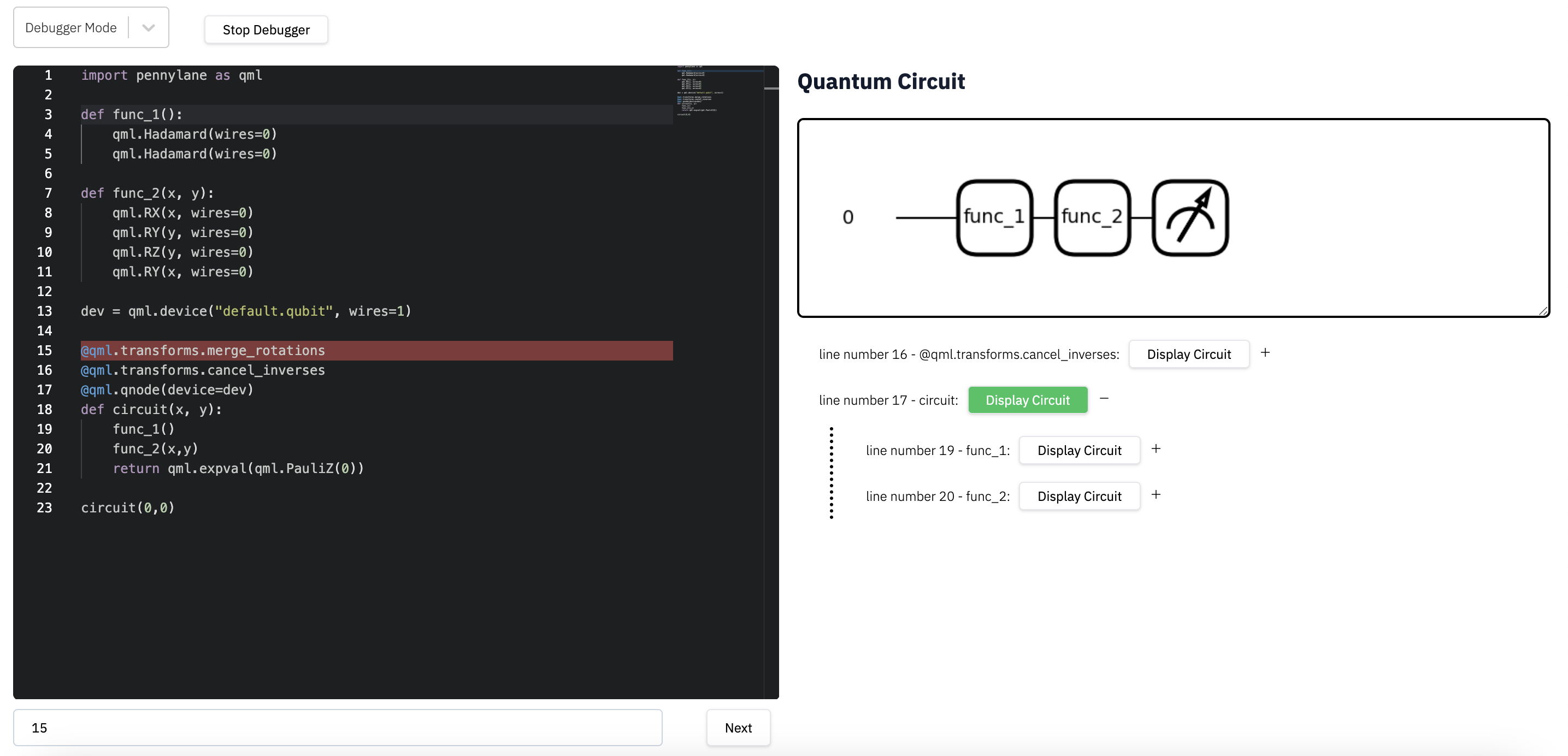}}
        \caption{
       Two transforms, \texttt{cancel\_inverses } and \texttt{merge\_rotations}, are being applied on the QNode \texttt{circuit}, with \texttt{cancel\_inverses }  applied first. Code execution has paused at line 15, and the tree structure displays the \texttt{cancel\_inverses} transform, indicating that it has been executed, above the QNode. A user can see the main function's output post-transform by clicking the transform name and visualize the quantum circuit by clicking the 'Display Circuit' button.}
        
        
        \label{fig:transform}
\end{figure*}

\toolname's debugger mode also supports transforms. Transforms alter the behavior of quantum functions and can be applied by decorating a QNode using the \texttt{@qml.transform} decorator~\cite{dimatteo2022quantum}. Transforms can be used for optimization and compilation, and these use cases are important enough that tools are being developed to debug them \cite{Timeline}. \toolname~allows users to set breakpoints precisely where transforms are applied. This functionality enhances a user's ability to visualize and comprehend the intricacies of transformation processes. \autoref{fig:transform} illustrates this capability.

\subsection{Real-time Development}
\subsubsection{Using the Real-time Development Mode}
Real-time development mode tracks code alterations and dynamically visualizes modifications. To begin real-time development, the user starts entering code in A1. React keeps track of what is typed in the code editor and displays an error message in A4 if there are syntax errors. 
Once a user executes a QNode, the code will execute in the Python backend and the appropriate quantum circuit visualization and tree structure are passed to the front end. All this happens in real time. When the user makes a change to their code, the above steps repeat.
\subsubsection{Benefits of the Real-time Development Mode}

Users can use real-time development to track output and circuit changes as they code. Continuous monitoring aids early issue detection such as unexpected outcomes and incorrect quantum operations.
Like the debugger mode, the function abstraction mirrors the actual code structure, ensuring that the quantum circuit reflects the ongoing development. This alignment aids in pinpointing discrepancies. Users can visualize quantum circuits for distinct subroutines and transforms, access function argument details, and analyze main function outputs within the tree component. Additionally, adjusting function arguments in real-time enables users to observe dynamic changes in the circuit structure, influenced by internal control flow or mid-circuit measurement outcomes. Furthermore, like the debugger mode, real-time development mode illustrates functions invoked within loops by depicting multiple instances in the circuit diagram and the tree diagram. 
This capability is crucial for identifying which function executions within the loops exhibit unexpected behavior, such as incorrect arguments or improper application of quantum operations.

\section{Plans for User Evaluation}
To evaluate the effectiveness of \toolname we plan to conduct quantitative and qualitative expert evaluations. We plan to recruit quantum computing graduate students and quantum software developers in the industry. These individuals will range from beginners to experts in the field of quantum computing and will have experience using Python and the PennyLane framework. This will ensure their ease in using\toolname and allow them to focus on debugging rather than PennyLane's intricacies. 

\subsubsection{Quantitative User Evaluation}
Quantitative evaluation will start with an overview of the study's objectives, emphasizing the motivation behind improving quantum debugging techniques and introducing the fundamental concepts integral to our approach. We will then transition into the training phase, in which we will demonstrate the capabilities of \toolname~on a well-known quantum algorithm typically taught in academic settings. Participants will be tasked with replicating these processes independently, allowing them to familiarize themselves with \toolname's functionalities.

After training participants will be presented with a quantum algorithm deliberately designed with bugs. Unlike the training phase, we will use an algorithm we participants are unlikely to know, but will provide them with pseudocode and specifications. Participants will then have 20 minutes to identify and rectify as many bugs as possible. \toolname~will be accessible via a website, and their interactions during this task will be recorded to analyze how they utilize \toolname, and assess their perspective on quantum bugs. 

\subsubsection{Qualitative User Evaluation}
The qualitative evaluation will occur after the quantitative evaluation. It will consist of a structured interview and a 5-point Likert Scale questionnaire.  We will conduct interviews with open-ended questions in which participants can share their experiences, challenges, preferences, and suggestions regarding \toolname. The focus will be to understand their interaction with the different usage modes, what they find valuable, and areas for improvement.

The questionnaire will cover aspects like ease of use, helpfulness of features such as real-time visualization and breakpoint setting, satisfaction levels with the debugger's functionality, the perceived value of specific features like tracking program execution, and the likelihood of recommending the debugger to others. 

\section{Conclusion and Future Work}
The challenges of quantum debugging highlight the urgent need for more research and development of tools. Effective quantum debugging requires visualization at different levels of abstraction in quantum programs, the ability to swiftly identify unexpected outputs, display executed functions and quantum operations, and respect control flow for accurate analysis,

This paper introduces \toolname, an interactive tool addressing these challenges. \toolname~has two modes, the debugger mode, and the real-time development mode. The debugger mode isolates quantum circuit components, monitors changes at breakpoints, observes algorithm inputs, and improves program comprehension. The real-time development mode tracks changes as well as visualizes them and enhances circuit representation in real time.

\toolname~has limitations that can be leveraged as opportunities for future work. Scalability presents a challenge, in particular for real-time development mode. Executing the circuit after each user change is resource-intensive, especially for complex quantum circuits. This leads to longer processing times and latency, affecting application responsiveness. To optimize this, we will monitor resource usage, performance metrics, and bottlenecks to identify inefficient code segments, optimize circuit execution, and dynamically adjust resource allocation. The debugger mode does not currently permit users to modify code while it's running; they must first stop the debugger to make changes. However, enabling code modifications without restarting the debugger can greatly accelerate development. Currently, all code must be uploaded in section A1, which works well for small quantum programs but becomes cumbersome for large programs that require multiple files. To address this, we plan to introduce a feature for uploading Python files and enabling debugging across all files.

Alongside addressing these limitations, we plan to add more features to \toolname. We will make the quantum circuit display interactive. Furthermore, we will add features that track user interaction with \toolname, providing valuable insights into user behavior and their perspectives on quantum bugs. Lastly, we will expand \toolname's capabilities by including additional information for functions such as quantum gate count and circuit depth. With these features added, \toolname~will be released open-source.




\section*{Acknowledgments}
MK acknowledges funding from the Natural Sciences and Engineering Research Council of Canada (NSERC) CREATE in Quantum Computing Program, grant number 543245, and the Four Year Doctoral Fellowship. Prashant Nair acknowledges funding from the NSERC Discovery Grant (RGPIN-2019-05059) and the National Research Council (NRC) Grant AQC 213-1. ODM acknowledges funding from NSERC (RGPIN-2022-04609, ALLRP 576833-22) and the Canada Research Chairs program. All authors thank the PennyLane team at Xanadu and members of the Quantum Software and Algorithms Research Lab at UBC for their feedback on \toolname, and Julia Rubin and Daria Ahrensmeier for useful discussions about user evaluations.

\newpage




%
\printbibliography

%









\end{document}